\documentclass[conference]{IEEEtran}

\usepackage{epsfig}
\usepackage{color}

\usepackage[cmex10]{amsmath}
\usepackage{amssymb}
\DeclareMathOperator*{\argmin}{argmin}
\DeclareMathOperator*{\argmax}{argmax}
\DeclareMathOperator*{\tr}{Tr}
\DeclareMathOperator*{\Cl}{Cl}
\DeclareMathOperator*{\Co}{Co}

\begin{document}
%
\title{Adaptive Sum Power Iterative Waterfilling for MIMO Cognitive Radio Channels}

\author{\IEEEauthorblockN{Rajiv Soundararajan and Sriram Vishwanath}
\IEEEauthorblockA{Department of Electrical and Computer Engineering, University of Texas at Austin\\
1 University Station C0803, Austin, TX 78712, USA\\
Email: soundara,sriram@ece.utexas.edu}}


%


\maketitle

\begin{abstract}
In this paper, the sum capacity of the Gaussian Multiple Input Multiple Output (MIMO) Cognitive Radio Channel (MCC) is expressed as a convex problem with finite number of linear constraints, allowing for polynomial time interior point techniques to find the solution. In addition, a specialized class of sum power iterative waterfilling algorithms is determined that exploits the inherent structure of the sum capacity problem. These algorithms not only determine the maximizing sum capacity value, but also the transmit policies that achieve this optimum. The paper concludes by providing numerical results which demonstrate that the algorithm takes very few iterations to converge to the optimum.
\end{abstract}


%
\IEEEpeerreviewmaketitle

\section{Introduction}
In recent years, the study of cognitive radios from an information theoretic perspective has gained prominence \cite{Jovicic}. As the Federal Communications Commission (FCC) determines the ways and bands in which cognitive radios can be used, it is imperative that we understand the fundamental limits of these radios to benchmark gains from their design and deployment. Cognitive radio channels refer to those media of communication in which cognitive radios operate, thereby efficiently utilizing available resources. Moreover, since most wireless systems these days use multiple antennas at the transmitter and receiver, it is important that we study the limits in a Multiple Input Multiple Output (MIMO) setting. The situation considered in this paper is different from the traditional class of MIMO problems on account of the intelligent and adaptive capabilities of the cognitive radios. \\

Based on the model proposed in \cite{Devroye2006}, the cognitive radio channel is an interference channel \cite{Carleial1978}\cite{Costa1985}\cite{Sato1981}\cite{Han1981} with degraded message sets in which the transmitter with a single message is called the ``primary" or ``licensed" user while the transmitter with both message sets is called the ``secondary" or ``cognitive" user. In this paper, we study the sum capacity of cognitive radios in a MIMO setting where both the primary and secondary transmitter and receivers have multiple antennas and the noise is Gaussian. The sum capacity enables the design of a MIMO cognitive radio system by specifying the sum rate required for the primary and the secondary users. Recently, an achievable region was found and shown to be optimal for the sum rate of the primary and secondary users \cite{Sridharan2007} under certain conditions on channel parameters. Though an achievable coding strategy based on Costa's dirty paper coding \cite{Costa1983} was shown to be optimal, an optimization over transmit covariances is required. The optimization to determine the sum capacity is in general a nonconvex problem and is hence computationally difficult \cite{Vishwanath2003a}.  \\

In this paper, we find that the nonconvex problem formulation is similar to a MIMO Broadcast Channel (BC) sum capacity problem formulation \cite{Vishwanath2003}. We therefore transform the nonconvex problem into a convex problem by using ``duality"' techniques as detailed in \cite{Vishwanath2003}. As a result we obtain a convex-concave game which can be solved in polynomial time. We propose efficient algorithms to find the saddle point of the problem and hence compute the sum capacity and optimal transmit policies. \\

The convex-concave game formulation of the sum capacity of the MIMO Cognitive Channel (MCC) is a minimax problem in which the inner maximization corresponds to computing the sum capacity of a MIMO Multiple Access Channel (MAC) subject to a sum power constraint. There are many efficient algorithms in literature that solve saddle point problems and they are analogous to convex optimization techniques like interior point and bundle methods \cite{Nemirovski2004}. However these algorithms are both much more involved than our algorithm and offer limited intuition about the structure of the optimal value. Our algorithm is based on the sum power iterative waterfilling algorithm for BC channels \cite{Jindal2005}, but is significantly different as the waterlevel is no longer given, but has to be discovered through {\it adaptation}. We thus call this strategy, the adaptive sum power waterfilling algorithm, to achieve our objective of solving the minimax problem.\\

The rest of the paper is organized as follows. In Section \ref{sec:system_model}, the system model is described and in Section \ref{sec:problem_statement}, we state the sum capacity problem for MCC. In Section \ref{sec:convex_problem}, the convex problem formulation is described. In Section \ref{sec:algorithm}, we propose the algorithm and numerical results are given in Section \ref{sec:numerical_results}. We conclude the paper in Section {\ref{sec:conclusion}. 


 



\section{System Model}\label{sec:system_model}
We use boldface letters to denote vectors and matrices. $\lvert \mathbf{H}\rvert$ denotes the determinant of the matrix $\mathbf{H}$ and $\tr(\mathbf{H})$ denotes the trace. For any general matrix $\mathbf{S}$, $\mathbf{S}^{\dag}$ denotes the conjugate transpose. $\mathbf{I}_n$ is the $n\times n$ identity matrix and $\mathbf{H}\succeq \mathbf{0}$ denotes that the square matrix $\mathbf{H}$ is positive semidefinite. If $\mathbf{E}$ is a set, then $\Cl(\mathbf{E})$ and $\Co(\mathbf{E})$ refer to the closure and convex hull of $\mathbf{E}$ respectively. \\

We consider the MCC illustrated in Fig. 1. The primary transmitter and receiver have $n_{p,t}$ and $n_{p,r}$ antennas while the cognitive transmitter and receiver have $n_{c,t}$ and $n_{c,r}$ antennas respectively. \\

\begin{figure}[!th]\label{fig_system_model}
\centering 
\scalebox{0.4}{
\input{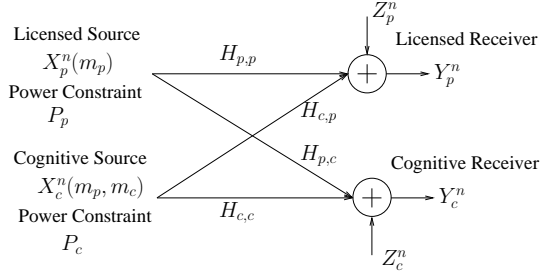}}
\caption{MIMO Cognitive Radio System Model}
\end{figure}



Let $\mathbf{X_p}(i) \in \mathbb{C}^{n_{p,t}\times 1}$ be the vector signal transmitted by the primary user and $\mathbf{X_c}(i) \in \mathbb{C}^{n_{c,t}\times 1}$ be the  vector signal transmitted by the cognitive user in time slot $i$. Let $\mathbf{H_{p,p}}$, $\mathbf{H_{p,c}}$, $\mathbf{H_{c,p}}$ and $\mathbf{H_{c,c}}$ be constant channel gain matrices as shown in Fig. 1. It is assumed that the licensed receiver knows $\mathbf{H_{p,p}}$ and $\mathbf{H_{c,p}}$, the licensed transmitter knows $\mathbf{H_{p,p}}$, the  cognitive transmitter knows $\mathbf{H_{p,c}}$, $\mathbf{H_{c,p}}$ and $\mathbf{H_{c,c}}$ and the cognitive receiver knows $\mathbf{H_{p,c}}$ and $\mathbf{H_{c,c}}$. These assumptions are made so as to make the problem tractable and thereby provide a benchmark on performance of the system. Let $\mathbf{Y_p}(i) \in \mathbb{C}^{n_{p,r}\times 1}$ and $\mathbf{Y_c}(i) \in \mathbb{C}^{n_{c,r}\times 1}$ be the signal received by the primary receiver and cognitive receiver respectively. The additive noise at the primary and secondary receivers are Gaussian, independent across time symbols and represented by $\mathbf{Z_p}(i) \in \mathbb{C}^{n_{p,r}\times 1}$ and $\mathbf{Z_c} \in \mathbb{C}^{n_{c,r}\times 1}$ respectively, where $\mathbf{Z_p}(i) \sim N(0,\mathbf{I}_{n_{p,r}})$ and $\mathbf{Z_c}(i) \sim N(0,\mathbf{I}_{n_{c,r}})$. $\mathbf{Z_p}(i)$ and $\mathbf{Z_c}(i)$ can be arbitrarily correlated between themselves. The received signal is mathematically represented as 

\begin{align*}
&\mathbf{Y_p}(i)=\mathbf{H_{p,p}X_{p}}(i)+\mathbf{H_{c,p}X_{c}}(i)+\mathbf{Z_{p}}(i)  \\
&\mathbf{Y_c}(i)=\mathbf{H_{p,c}X_{p}}(i)+\mathbf{H_{c,c}X_{c}}(i)+\mathbf{Z_{c}}(i).
\end{align*}

The covariance matrices of the primary and cognitive input signals are $\mathbf{\Sigma_{p}}(i)$ and $\mathbf{\Sigma_{c}}(i)$. The primary and secondary transmitters are subject to average power constraints $P_p$ and $P_c$ respectively. Thus, 
\begin{align*}
&\sum_{i=1}^{n}\tr(\mathbf{\Sigma_{p}}(i)) \leq nP_p \\
&\sum_{i=1}^{n}\tr(\mathbf{\Sigma_{c}}(i)) \leq nP_c.
\end{align*}
   
\section{Problem Statement}\label{sec:problem_statement}
In this section, we restate the sum capacity of the MCC as in \cite{Sridharan2007}. Before doing so, we develop the required notation for the same.  
Let $\mathbf{G}=[\mathbf{H_{p,p}} \quad \mathbf{H_{c,p}}]$. The set $\mathcal{R}_{ach}$ is defined as 
\begin{eqnarray*}
\mathcal{R}_{ach} = \left\{ \begin{array}{l}\Big((R_p,R_c),\mathbf{\Sigma_{p}},\mathbf{\Sigma_{c,p}},\mathbf{\Sigma_{c,c}},\mathbf{Q}\Big) : \\
\quad R_p, R_c \geq 0, \mathbf{\Sigma_{p}},\mathbf{\Sigma_{c,p}},\mathbf{\Sigma_{c,c}} \succeq \mathbf{0} \\
\quad R_p  \leq  \log(\frac{\lvert \mathbf{I}+\mathbf{G\Sigma_{p,net}G^{\dag}}+\mathbf{H_{c,p}\Sigma_{c,c}} \mathbf{H}_{\mathbf{c,p}}^{\dag}\rvert}{\lvert \mathbf{I}+\mathbf{H_{c,p}\Sigma_{c,c}} \mathbf{H}_{\mathbf{c,p}}^{\dag}\rvert})\\
\quad R_c \leq \log(\lvert \mathbf{I}+\mathbf{H_{c,c}\Sigma_{c,c}} \mathbf{H}_{\mathbf{c,c}}^{\dag}\rvert)\\
\quad \mathbf{\Sigma_{p,net}}= \left( \begin{array}{ccc} 
\mathbf{\Sigma_{p}} & \mathbf{Q}   \\
\mathbf{Q}^{\dag} &  \mathbf{\Sigma_{c,p}} 
\end{array} \right),\\
\quad \tr(\mathbf{\Sigma_{p}}) \leq P_p,  \tr(\mathbf{\Sigma_{c,p}}+\mathbf{\Sigma_{c,c}}) \leq P_c
\end{array} \right\}
\end{eqnarray*} where $\mathbf{\Sigma_{p, net}}$ is a $(n_{p,t} + n_{c,t}) \times (n_{p,t} + n_{c,t})$ covariance matrix while $\mathbf{\Sigma_{c,c}}$ is a $n_{c,t} \times n_{c,t}$ covariance matrix. $\mathbf{\Sigma_p}$ and $\mathbf{\Sigma_{c,p}}$ are principal submatrices of $\mathbf{\Sigma_{p, net}}$ of dimensions $n_{p,t} \times n_{p,t}$ and $n_{c,t} \times n_{c,t}$ respectively. The set of all rate pairs $\mathcal{R}_{in}$ is given by
\begin{align*}\label{eqn:achievable_region} 
\mathcal{R}_{in}=&\Cl\Bigg(\Co\Bigg\{(R_p, R_c) : \exists\ \mathbf{\Sigma_p},
\mathbf{\Sigma_{c,p}}, \mathbf{\Sigma_{c,c}} \succeq \mathbf{0} \ \textrm{and} \nonumber \\ 
&\mathbf{Q} : \bigg((R_p, R_c), \mathbf{\Sigma_p}, \mathbf{\Sigma_{c,p}}, \mathbf{\Sigma_{c,c}}, \mathbf{Q}\bigg) \in
\mathcal{R}_{ach}\Bigg\}\Bigg).
\end{align*}
It is shown in \cite{Sridharan2007} that $\mathcal{R}_{in}$ is an inner bound on the capacity region of the MCC. \\

Let $\mathbf{G}_{\alpha}=[\mathbf{H_{p,p}} \quad \frac{\mathbf{H_{c,p}}}{\sqrt{\alpha}}]$ and $\mathbf{K}_{\alpha} = [\mathbf{0} \quad \frac{\mathbf{H_{c,c}}}{\sqrt{\alpha}}]$. The set $\mathcal{R}_{part}^\alpha$ is defined as

\begin{eqnarray*}
\mathcal{R}_{part}^{\alpha}=\left\{
\begin{array}{l}
\Big((R_p,R_c),\mathbf{Q_{p}},\mathbf{\Sigma_{c,c}},\mathbf{Q}\Big): \\
\quad R_p \geq 0, R_c \geq 0, \mathbf{Q_{p}}, \mathbf{\Sigma_{c,c}}\succeq \mathbf{0}  \\
\quad R_p  \leq  \log(\frac{\lvert \mathbf{I}+\mathbf{G}_{\alpha}\mathbf{Q_{p}}\mathbf{G}_{\alpha}^{\dag}+\frac{1}{\alpha}\mathbf{H_{c,p}\Sigma_{c,c}} \mathbf{H}_{\mathbf{c,p}}^{\dag}\rvert}{\lvert \mathbf{I}+\frac{1}{\alpha}\mathbf{H_{c,p}\Sigma_{c,c}} \mathbf{H}_{\mathbf{c,p}}^{\dag}\rvert})  \\
\quad R_c \leq \log(\lvert \mathbf{I}+\frac{1}{\alpha}\mathbf{H_{c,c}\Sigma_{c,c}} \mathbf{H}_{\mathbf{c,c}}^{\dag}\rvert)  \\
\quad \tr(\mathbf{Q_{p}}) +  \tr(\mathbf{\Sigma_{c,c}}) \leq P_p+\alpha P_c 
\end{array} \right\}
\end{eqnarray*} and the set $\mathcal{R}_{part, out}^{\alpha}$ is defined as
\begin{eqnarray*}\label{eqn:partial_converse_region}
\begin{array}{lcr}
\mathcal{R}_{part, out}^{\alpha} = \Cl\Bigg(\Co\Bigg\{(R_p, R_c) : \exists
\mathbf{Q_p}, \mathbf{\Sigma_{c,c}} \succeq \mathbf{0} \\
\quad \quad \quad \quad \quad \quad \textrm{such that } ((R_p, R_c), \mathbf{Q_p}, \mathbf{\Sigma_{c,c}}) \in
\mathcal{R}_{part}^{\alpha}\Bigg\}\Bigg). 
\end{array}
\end{eqnarray*}
It is also shown in \cite{Sridharan2007} that $\mathcal{R}_{part,out}^{\alpha}$ is an outer bound on the capacity region that includes the sum rate when certain conditions that depend on the channel parameters are satisfied. We compute the sum capacity of the MCC under those conditions.\\

We now restate Theorem 3.3 from \cite{Sridharan2007} which paves the way for the MCC sum capacity problem formulation. For any $\mu \geq 1$,
\begin{displaymath}
\max_{(R_p, R_c) \in \mathcal{R}_{in}} \mu R_p + R_c = \inf_{\alpha
> 0} \max_{(R_p, R_c) \in \mathcal{R}_{part, out}^{\alpha}} \mu R_p + R_c.
\end{displaymath}
Therefore, the sum capacity of the MCC (denoted by $\mathcal{C}_{MCC}(\mathbf{G_{\alpha}},\mathbf{H_{c,c}})$) is expressed as 
\begin{equation}\label{eqn:sum_capacity}
\mathcal{C}_{MCC}(\mathbf{G}_{\alpha},\mathbf{H_{c,c}}) = \inf_{\alpha
> 0} \max_{(R_p, R_c) \in \mathcal{R}_{part, out}^{\alpha}} R_p + R_c.
\end{equation}

\section{Formulation as a Convex Problem}\label{sec:convex_problem}
The inner maximization in the sum capacity of the MCC, stated above, corresponds to computing the sum capacity of a degraded broadcast channel in which the transmitters cooperate with a sum power constraint. This is illustrated in Fig. 2. We observe from the expression for the sum capacity in (\ref{eqn:sum_capacity}) that the inner maximization problem is not a concave function of the covariance matrices $\mathbf{Q_{p}}$ and $\mathbf{\Sigma_{c,c}}$. Thus it is difficult to solve the entire problem using numerical techniques. However, as in \cite{Jindal2005}, we can use ``duality" to transform the inner maximization problem into a sum capacity problem for the MAC with the same sum power constraint. This can be done because it is shown in \cite{Vishwanath2003} that sum capacity of the BC is exactly equal to the sum capacity of the dual MAC. These results enable us to convert the original problem to a convex-concave game.  \\


\begin{figure}[!th]
\centering
\scalebox{0.4}{
\input{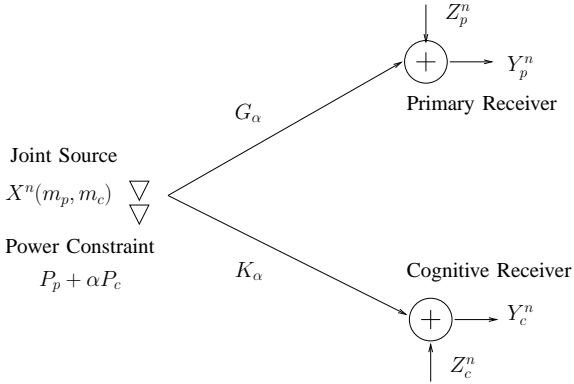}}
\caption{MIMO Broadcast Channel}
\end{figure}

Let $\mathbf{Q_{c}}=\left[\begin{array}{ccc}
\mathbf{0} & \mathbf{0} \\
\mathbf{0} & \mathbf{\Sigma_{c,c}}\end{array}\right]$, where the zero matrices have appropriate dimensions such that $\mathbf{\Sigma_{c,c}}$ has dimension $n_{c,t}\times n_{c,t}$ and $\mathbf{Q_{c}}$ has dimension $(n_{p,t}+n_{c,t})\times(n_{p,t}+n_{c,t})$. The sum capacity of the MCC can therefore be expressed as  
\begin{align}\label{eqn:mcc_sum_capacity}
&\mathcal{C}_{MCC} \nonumber = \inf_{\alpha > 0} \max_{\mathbf{S}_1,\mathbf{S}_2} \log\lvert\mathbf{I}+\mathbf{G}_{\alpha}^{\dag}\mathbf{S}_1\mathbf{G}_{\alpha}+\mathbf{K}_{\alpha}^{\dag}\mathbf{S}_2 \mathbf{K}_{\alpha}\rvert \nonumber \\
& \textrm{subject to} \ \ \mathbf{S}_1,\mathbf{S}_2 \succeq 0, \tr(\mathbf{S}_1)+\tr(\mathbf{S}_2) \leq P_p + \alpha P_c 
\end{align}
where the maximization is performed over covariance matrices $\mathbf{S}_{1}$ and $\mathbf{S}_{2}$, which are obtained using BC-to-MAC transformations of $\mathbf{Q_{p}}$ and $\mathbf{Q_{c}}$ such that the sum of the traces of $\mathbf{S}_{1}$ and $\mathbf{S}_{2}$ satisfies the sum power constraint $P_p+\alpha P_c$. This new problem is concave in the covariance matrices $\mathbf{S}_{1}$ and $\mathbf{S}_{2}$ and convex in the scalar $\alpha$ with linear power constraints and is thus a convex-concave game. The proof of the fact that the problem is convex in $\alpha$ is given in the appendix. This min-max problem can be solved by using interior point methods or other equivalent convex optimization methods for saddle point problems which have polynomial time complexity \cite{Nemirovski2004}. Once the optimal $\mathbf{S}_{1}$ and $\mathbf{S}_{2}$ are obtained, we can apply the MAC-to-BC transformation \cite{Vishwanath2003} to obtain the optimal transmit policies of our original problem. The MAC-to-BC transformation, takes a set of MAC covariance matrices  and outputs a set of BC covariance matrices which achieve the same sum rate as the MAC covariance matrices. 

\section{Adaptive Sum Power Iterative Water-filling}\label{sec:algorithm}
We propose a class of algorithms to compute the sum capacity of the MCC. These algorithms are motivated by the sum power water-filling algorithms developed for computing the sum capacity and obtaining the optimal transmit policies for the MIMO BC. As in \cite{Jindal2005}, we obtain a dual convex problem corresponding to a MAC. 
This dual MAC problem is a convex problem and thus can be solved using the multitude of convex solvers in polynomial time. Our intention, as that in \cite{Vishwanath2003a}\cite{Yu2004} is to exploit the problem's structure to yield a more intuitive and easy-to-implement algorithm for this problem. Specifically, it is to derive an algorithm that reflects and converges to the KKT conditions corresponding to cognitive radio sum capacity.\\

Individual power iterative waterfilling was found to achieve the capacity of a MAC channel with separate power constraints per user in \cite{Vishwanath2003a}. This was then extended to BCs by the sum power iterative waterfilling algorithm in \cite{Yu2004}, which is based on the dual MAC's KKT conditions. While neither of these two algorithms works directly for the MIMO cognitive radio channel, we use analogous principles to derive an adaptive sum power iterative algorithm. \\

Before proceeding further, we review the waterfilling algorithm for the single user point to point MIMO case. Consider the problem, 
\begin{equation}\label{eqn:mimo}
\max_{\tr(\mathbf{S}) \leq P, \ \mathbf{S} \succeq 0} \frac{1}{2}\log\lvert\mathbf{I}+\mathbf{H}^{\dag}\mathbf{S}\mathbf{H}\rvert.
\end{equation}
Let the eigen values of the covariance matrix $\mathbf{S}$ of size $n \times n$ be $\lambda_1,\lambda_2,\ldots,\lambda_n$ and the singular values of the channel matrix $\mathbf{H}$ be $\sigma_1,\sigma_2,\ldots,\sigma_n$. Then the waterfilling solution is given by 
\begin{eqnarray}
\lambda_{i} + 1/\sigma_{i}^{2}=K, &\textrm{if}& 1/\sigma_{i}^{2} < K \\
\lambda_{i} = 0, &\textrm{if}&  1/\sigma_{i}^{2} \geq K,
\end{eqnarray}
where K is a constant such that $\sum_{i}\lambda_{i}=P$. KKT conditions, along with complementary slackness yield conditions (4) and (5) \cite{Telatar1999}\cite{Goldsmith2003}. We now recall the sum power iterative waterfilling solution for the MIMO MAC as studied in \cite{Yu2004}. The problem is stated below:
\begin{equation}
\max_{\tr(\sum\mathbf{S}_{i}) \leq P, \ \mathbf{S}_{i} \succeq 0} \frac{1}{2}\log\lvert\mathbf{I}+\sum_{i}\mathbf{H}_{i}^{\dag}\mathbf{S}_{i}\mathbf{H}_{i}\rvert
\end{equation}
The KKT conditions for this problem are similar to the point to point case except that in the MAC case, the channel H in (\ref{eqn:mimo}) for User $i$ is replaced by the effective channel matrix $\mathbf{H}_{i,eff} = \mathbf{H}_{i}(\mathbf{I}+\sum_{j\ne i}\mathbf{H}_{j}^{\dag}\mathbf{S}_{j}\mathbf{H}_{j})$. In the sum power waterfilling algorithm, we waterfill for all the users simultaneously, while in individual power iterative waterfilling, user waterfilling is sequenced \cite{Jindal2005}.\\

The problem considered in this paper has a min-max formulation as given in (\ref{eqn:mcc_sum_capacity}) unlike the pure max formulations for the MAC and BC capacity problems \cite{Jindal2005}\cite{Yu2004}. The idea behind our algorithm is to start with a feasible choice for $\alpha$. For a given $\alpha$, the joint water-filling on the two users is the optimal strategy for the inner maximization problem. In (\ref{eqn:mcc_sum_capacity}), we perform a joint water-fill exactly once for the inner maximization, use this solution to solve the outer minimization with respect to $\alpha$ and iterate between the two.  Thus, we end up with one maximization problem (for a choice of $\alpha$) given by
\begin{align}
&\mathbf{S}_1,\mathbf{S}_2=\argmax_{\mathbf{T}_1,\mathbf{T}_2} \log\lvert\mathbf{I}+\mathbf{G}_{\alpha}^{\dag}\mathbf{T}_1\mathbf{G}_{\alpha}+\mathbf{K}_{\alpha}^{\dag}\mathbf{T}_2 \mathbf{K}_{\alpha}\rvert \nonumber \\
&\textrm{subject to} \ \ \mathbf{T}_1,\mathbf{T}_2 \succeq 0, \ \ \tr(\mathbf{T}_1)+\tr(\mathbf{T}_1) \leq P_p + \alpha P_c \nonumber
\end{align}
and a minimization problem (for a given choice of covariances) as:
\begin{align}
&\alpha = \argmin_{\beta > 0} \log\lvert\mathbf{I}+\mathbf{G}_{\beta}^{\dag}\gamma\mathbf{S}_1\mathbf{G}_{\beta}+\mathbf{K}_{\beta}^{\dag}\gamma\mathbf{S}_2 \mathbf{K}_{\beta}\rvert \nonumber \\
& \textrm{where } \gamma = \frac{P_p+\beta P_c}{P_p+\alpha^{(n-1)}P_c} \textrm{ is a scalar} \nonumber \\
& \textrm{ and } \alpha^{(n-1)} \textrm{ is the previous iterate of } \alpha. \nonumber
\end{align}
The solution of one is fed to the other, and the process is repeated until convergence. This forms the core of the adaptive sum power iterative waterfilling algorithm.\\

We refer to the procedure detailed above as Algorithm 1. As the minimization with respect to $\alpha$ can be somewhat involved (even though $\alpha$ is a scalar), we construct another algorithm we call Algorithm 2. In Algorithm 2, we only obtain a descent at each iteration by a simple line search like Newton search. We do not solve the outer minimization problem completely at every iteration as in Algorithm 1, which further simplifies the overall algorithm. In Section \ref{sec:numerical_results}, we illustrate through examples that this highly simplified algorithm, with one waterfill and one descent in each iteration has nearly as good a convergence rate as exact solutions at each step.  In the following, $n$ refers to the iteration number.\\
\\
Main Algorithm (Algorithm 1):
\begin{enumerate}
\item Initialize $\alpha^{(0)}$ to any number greater than 0. \\
\item Generate the effective channels as 
\begin{align}
\mathbf{G}_{\alpha,eff}^{(n)}&=&\mathbf{G}_{\alpha^{(n-1)}}\big(\mathbf{I}+\mathbf{K}_{\alpha^{(n-1)}}^{\dag}\mathbf{S}_2^{(n-1)} \mathbf{K}_{\alpha^{(n-1)}}\big)^{-1/2} \nonumber \\
\cr \mathbf{K}_{{\alpha},eff}^{(n)}&=&\mathbf{K}_{\alpha^{(n-1)}}\big(\mathbf{I}+\mathbf{G}_{\alpha^{(n-1)}}^{\dag}\mathbf{S}_1^{(n-1)}\mathbf{G}_{\alpha^{(n-1)}}\big)^{-1/2}. \nonumber
\end{align}
\item Obtain covariance matrices $\mathbf{S}_1^{(n)}$ and $\mathbf{S}_2^{(n)}$ by performing a joint waterfill with power $P_p+\alpha^{(n-1)} P_c$. 
\begin{align*}
&\{\mathbf{S}_1^{(n)},\mathbf{S}_2^{(n)}\} =\argmax_{\mathbf{T}_1,\mathbf{T}_2} \log\lvert\mathbf{I}+\mathbf{G}_{\alpha^{(n-1)}}^{\dag}\mathbf{T}_1\mathbf{G}_{\alpha^{(n-1)}} \\ 
& \quad \quad \quad \quad \quad \quad \quad \quad \quad \quad \quad \quad \quad +\mathbf{K}_{\alpha^{(n-1)}}^{\dag}\mathbf{T}_2 \mathbf{K}_{\alpha^{(n-1)}}\rvert   \\
&\textrm{subject to} \ \mathbf{T}_1,\mathbf{T}_2 \succeq 0 \textrm{ and} \\
&\tr(\mathbf{T}_1)+\tr(\mathbf{T}_1) \leq P_p + \alpha^{(n-1)} P_c. 
\end{align*}
\item Use $\mathbf{S}_1^{(n)}$ and $\mathbf{S}_2^{(n)}$ from Step 3 to obtain $\alpha^{(n)}$ by solving the following univariate optimization problem.
\begin{align*}
&\alpha^{(n)} = \argmin_{\beta > 0} \log\lvert\mathbf{I}+\mathbf{G}_{\beta}^{\dag}\gamma\mathbf{S}_1^{(n)}\mathbf{G}_{\beta}+\mathbf{K}_{\beta}^{\dag}\gamma\mathbf{S}_2^{(n)} \mathbf{K}_{\beta}\rvert \\
& \textrm{where } \gamma = \frac{P_p+\beta P_c}{P_p+\alpha^{(n-1)}P_c} \textrm{ is a scalar.} \nonumber
\end{align*}

\item Return to Step 2 until parameters converge. 
\end{enumerate}

As mentioned before, the above algorithm is a very intuitive extension to the sum power waterfilling algorithm for the MIMO BC channel. The intuition arises from the fact that at the saddle point, the KKT conditions must be satisfied for both the max and the min problems. Although we may not be able to always guarantee convergence of the algorithm to the optimal solution, when it does converge, the algorithm takes very few iterations to do so.

\section{Numerical Results}\label{sec:numerical_results}
In this section we present numerical results to compare the behavior of Algorithms 1 and 2. In Fig. 3, we plot the sum rate versus the number of iterations for a MCC with two antennas at both the primary and cognitive receiver and one antenna each at the primary and cognitive transmitter with power constraints $P_{p} = P_{c} = 5$. The channel matrices are 
\begin{align*}
&\mathbf{G}_{\alpha} = \left[\begin{array}{ccc}
-0.4326  & -1.6656 \\
0.1253   &  0.2877\end{array}\right]  \\
\textrm{and }&\mathbf{K}_{\alpha} = \left[\begin{array}{ccc}
0  & -1.1465 \\
0  & 1.1909\end{array}\right]
\end{align*} 
when $\alpha=1$. We find that both Algorithm 1 and Algorithm 2 converge to the same sum rate. However this may not always happen depending on the initial conditions chosen for the Newton search. We also observe that in some cases Algorithm 1 converges in fewer iterations when compared to Algorithm 2. 

\begin{figure}[!t]
\centering 
\includegraphics[width=3.5in]{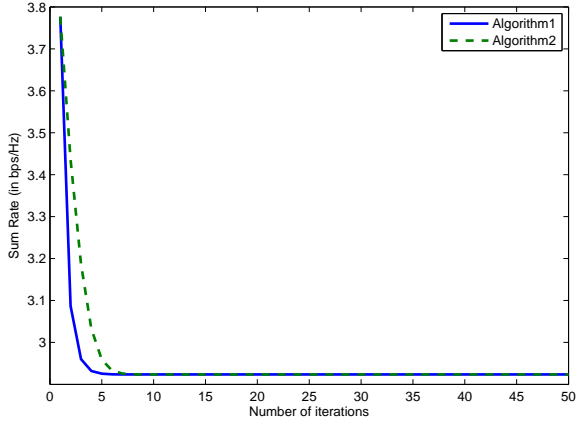}
\caption{Convergence of Algorithm 1 and Algorithm 2 to the sum rate}
\label{fig_sim}
\end{figure}


\section{Conclusion} \label{sec:conclusion}
In this paper, we proposed a class of algorithms to compute the sum capacity and the optimal transmit policies of the MCC. This was made possible by transforming the MCC sum capacity problem as a convex problem using MAC-BC (otherwise called as uplink-downlink) duality. The algorithm performs a sum power waterfill at each iteration, while simultaneously adapting the waterlevel at each iteration.

\appendix
\textit{Proof of convexity of the objective in (\ref{eqn:mcc_sum_capacity}) in $\alpha$}: Let 
$F(\alpha) = \log\lvert\mathbf{I}+\mathbf{G}_{\alpha}^{\dag}\mathbf{S}_1\mathbf{G}_{\alpha}+\mathbf{K}_{\alpha}^{\dag}\mathbf{S}_2 \mathbf{K}_{\alpha} \rvert$. Hence, 

\begin{align*}
F =& \log \Bigg\lvert \begin{array}{ccc}
\mathbf{I}+\mathbf{H}_{\mathbf{p,p}}^{\dag}\mathbf{S}_1\mathbf{H}_{\mathbf{p,p}} & \frac{\mathbf{H}_{\mathbf{p,p}}^{\dag}\mathbf{S}_1\mathbf{H}_{\mathbf{c,p}}}{\sqrt{\alpha}}  \\
\frac{\mathbf{H}_{\mathbf{c,p}}^{\dag}\mathbf{S}_1\mathbf{H}_{\mathbf{p,p}}}{\sqrt{\alpha}} & \mathbf{I}+\frac{\mathbf{H}_{\mathbf{c,p}}^{\dag}\mathbf{S}_1\mathbf{H}_{\mathbf{c,p}}+\mathbf{H}_{\mathbf{c,c}}^{\dag}\mathbf{S}_2\mathbf{H}_{\mathbf{c,c}}}{\alpha} 
\end{array} \Bigg\rvert \\
=&\log \lvert\mathbf{I}+\mathbf{H}_{\mathbf{p,p}}^{\dag}\mathbf{S}_1\mathbf{H}_{\mathbf{p,p}}\rvert \\
&+\log \Bigg\lvert \mathbf{I} +  \frac{\mathbf{H}_{\mathbf{c,p}}^{\dag}\mathbf{S}_1\mathbf{H}_{\mathbf{c,p}}+\mathbf{H}_{\mathbf{c,c}}^{\dag}\mathbf{S}_2\mathbf{H}_{\mathbf{c,c}}}{\alpha} \\ & \quad   -\frac{\mathbf{H}_{\mathbf{c,p}}^{\dag}\mathbf{S}_1\mathbf{H}_{\mathbf{p,p}}(\mathbf{I}+\mathbf{H}_{\mathbf{p,p}}^{\dag}\mathbf{S}_1\mathbf{H}_{\mathbf{p,p}})^{-1}\mathbf{H}_{\mathbf{p,p}}^{\dag}\mathbf{S}_1\mathbf{H}_{\mathbf{c,p}}}{\alpha} \Bigg\rvert.
\end{align*}
Thus $F$ is of the form $F = c + \log\Big\lvert \mathbf{I} + \frac{\mathbf{A}}{\alpha}\Big \rvert$ where $c=\log \lvert\mathbf{I}+\mathbf{H}_{\mathbf{p,p}}^{\dag}\mathbf{S}_1\mathbf{H}_{\mathbf{p,p}}\rvert$ and $\mathbf{A} = \mathbf{H}_{\mathbf{c,p}}^{\dag}\mathbf{S}_1\mathbf{H}_{\mathbf{c,p}}+\mathbf{H}_{\mathbf{c,c}}^{\dag}\mathbf{S}_2\mathbf{H}_{\mathbf{c,c}} -\mathbf{H}_{\mathbf{c,p}}^{\dag}\mathbf{S}_1\mathbf{H}_{\mathbf{p,p}}(\mathbf{I}+\mathbf{H}_{\mathbf{p,p}}^{\dag}\mathbf{S}_1\mathbf{H}_{\mathbf{p,p}})^{-1}\mathbf{H}_{\mathbf{p,p}}^{\dag}\mathbf{S}_1\mathbf{H}_{\mathbf{c,p}}$. From matrix theory \cite[Chap. 7]{Horn}, we know that $\mathbf{I} + \frac{\mathbf{A}}{\alpha}$ is positive semidefinite for all $\alpha>0$ since $\mathbf{I}+\mathbf{G}_{\alpha}^{\dag}\mathbf{S}_1\mathbf{G}_{\alpha}+\mathbf{K}_{\alpha}^{\dag}\mathbf{S}_2 \mathbf{K}_{\alpha}$ and $\mathbf{I}+\mathbf{H}_{\mathbf{p,p}}^{\dag}\mathbf{S}_1\mathbf{H}_{\mathbf{p,p}}$ are positive semidefinite for all $\alpha>0$. Let $\lambda_{i}, i=1,2,3,\ldots, n_{c,t}$ be the eigenvalues of $\mathbf{A}$. Therefore, for every $i$, $1+\frac{\lambda_{i}}{\alpha} \geq 0$ for all $\alpha>0$ which implies $\lambda_{i} \geq -\alpha$ for all $\alpha>0$. Hence $\lambda_{i} \geq 0$ for $i=1,2,3,\ldots, n_{c,t}$ and $\mathbf{A}$ is positive semidefinite. $\frac{\partial^{2}F}{\partial {\alpha}^{2}}$ is given by 
\begin{align*}
\frac{\partial^{2}F}{\partial {\alpha}^{2}} = &\tr\Bigg[\frac{2\mathbf{A}}{\alpha^{3}}\Big(\mathbf{I}+\frac{\mathbf{A}}{\alpha}\Big)^{-1}\Bigg] \\
&-\tr\Bigg[\frac{\mathbf{A}}{\alpha^{2}}\Big(\mathbf{I}+\frac{\mathbf{A}}{\alpha}\Big)^{-1}\frac{\mathbf{A}}{\alpha^{2}}\Big(\mathbf{I}+\frac{\mathbf{A}}{\alpha}\Big)^{-1}\Bigg]\\
= &\tr\Bigg[\frac{\mathbf{A}}{\alpha^{3}}\Big(\mathbf{I}+\frac{\mathbf{A}}{\alpha}\Big)^{-1}\Big(2\mathbf{I}+\frac{\mathbf{A}}{\alpha}\Big)\Big(\mathbf{I}+\frac{\mathbf{A}}{\alpha}\Big)^{-1}\Bigg].
\end{align*}
Using matrix theory results \cite[Chap. 7]{Horn} we can further show that $\frac{\partial^{2}F}{\partial {\alpha}^{2}}\geq 0$ for all $\alpha > 0$. Thus $F(\alpha)$ is convex in $\alpha$.


\section*{Acknowledgment}
This work was supported in part by grants from THECB-ARP and ARO YIP.
%



%
%
%



\end{document}